\documentclass[conference,10pt,top=0.7in,bottom=1in]{IEEEtran}
\IEEEoverridecommandlockouts

\usepackage{color}
\usepackage{cite}
\usepackage{amsmath,amssymb,amsfonts}

\usepackage{graphicx}
\usepackage{textcomp}
\usepackage{xcolor}
\usepackage[algo2e]{algorithm2e}
\usepackage{algorithm}

\def\BibTeX{{\rm B\kern-.05em{\sc i\kern-.025em b}\kern-.08em
    T\kern-.1667em\lower.7ex\hbox{E}\kern-.125emX}}
\makeatletter
\newcommand\fs@spaceruled{\def\@fs@cfont{\bfseries}\let\@fs@capt\floatc@ruled
  \def\@fs@pre{\vspace{0.5\baselineskip}\hrule height.8pt depth0pt \kern2pt}%
  \def\@fs@post{\kern1pt\hrule\relax}%
  \def\@fs@mid{\kern2pt\hrule\kern2pt}%
  \let\@fs@iftopcapt\iftrue}
  \newcommand\fs@betterruled{%
  \def\@fs@cfont{\bfseries}\let\@fs@capt\floatc@ruled
  \def\@fs@pre{\vspace*{5pt}\hrule height.8pt depth0pt \kern2pt}%
  \def\@fs@post{\kern2pt\hrule\relax}%
  \def\@fs@mid{\kern2pt\hrule\kern2pt}%
  \let\@fs@iftopcapt\iftrue}
\floatstyle{betterruled}
\restylefloat{algorithm}
\makeatother
    
\begin{document}

\title{ Flexible Semantic-Aware Resource Allocation: Serving More Users Through Similarity Range Constraints
}\vspace{-4mm}
\author{
\IEEEauthorblockN{\fontsize{10}{10}\selectfont Nasrin Gholami\IEEEauthorrefmark{1}, Neda Moghim\IEEEauthorrefmark{1}\IEEEauthorrefmark{2}, Behrouz Shahgholi Ghahfarokhi\IEEEauthorrefmark{1}, Pouyan Salavati\IEEEauthorrefmark{1}, \\ Christo Kurisummoottil Thomas\IEEEauthorrefmark{3}, Sachin Shetty\IEEEauthorrefmark{2}, Tahereh Rahmati\IEEEauthorrefmark{1}}
\IEEEauthorrefmark{1}Department of Computer Engineering,
University of Isfahan,\\ \IEEEauthorrefmark{2}Center for Secure and Intelligent Critical Systems,
Old Dominion University, VA, US \\
\IEEEauthorrefmark{3}
Bradley Department of Electrical and Computer Engineering,
Virginia Tech, Arlington,
VA, USA \\
Emails: \{ngholami1234, n.moghim, shahgholi, pouyan\}@eng.ui.ac.ir, \\ \fontsize{8}{8} christokt@vt.edu, sshetty@odu.edu, tahereh.rahmati@gmail.com\vspace{-5mm}}


\maketitle

\begin{abstract}
Semantic communication (SemCom) aims to enhance the resource efficiency of next-generation networks by transmitting the underlying meaning of messages, focusing on information relevant to the end user. Existing literature on SemCom primarily emphasizes learning the encoder and decoder through end-to-end deep learning frameworks, with the objective of minimizing a task-specific semantic loss function. Beyond its influence on the physical  and application layer design, semantic variability across users in multi-user systems enables the design of resource allocation schemes that incorporate user-specific semantic requirements. To this end, \emph{a semantic-aware resource allocation} scheme is proposed with the objective of maximizing transmission and semantic reliability, ultimately increasing the number of users whose semantic requirements are met. The resulting resource allocation problem is a non-convex mixed-integer nonlinear program (MINLP), which is known to be NP-hard. To make the problem tractable, it is decomposed into a set of sub-problems, each of which is efficiently solved via geometric programming techniques. Finally, simulations demonstrate that the proposed method improves user satisfaction by up to $17.1\%$ compared to state of the art methods based on quality of experience-aware SemCom methods.
\end{abstract}


\section{Introduction}
Semantic communication (SemCom) has recently gained prominence as a promising approach for managing the growing data demands of connected intelligent systems such as augmented reality/virtual reality (AR/VR), autonomous driving, and intelligent transportation systems, all of which require high data throughput and ultra-low latency. Unlike conventional communication systems, which focus on transmitting symbols or bits, SemCom prioritizes the meaning of the transmitted data in the communication context. By transmitting only relevant information, SemCom improves resource efficiency and network capacity. Artificial intelligence (AI) techniques are used to extract semantic information from the source data, encode it and reconstruct the erroneous channel output at the receiver \cite{b1,b2,b3}. While significant progress has been made, key challenges remain, particularly in optimizing communication resources in a way that is sensitive to the semantic relevance of transmitted content. Addressing this semantic-aware resource allocation is essential for advancing the efficiency and reliability of next-generation networks. 

The benefits of semantic-aware resource allocation vary depending on the end-user’s intent, which may range from complete data reconstruction, referred to as semantic-oriented communication (SOC), to achieving task-specific objectives, known as task- or goal-oriented semantic communication.  In this context, reconstruction performance at the receiver can be quantified using \emph{semantic similarity}, which measures the extent to which the recovered data aligns with the intended communication objective. This measure depends on both the amount of transmitted semantic information, determined by the compression rate, and the signal-to-noise ratio (SNR) of the communication channel. While traditional networks primarily focus on communication reliability, it is increasingly critical to design novel resource allocation schemes that also account for semantic similarity in order to optimize end-to-end performance. 

\vspace{-2mm}\subsection{Related works}

Several studies \cite{b1,b4,b5,b6,b7,b8,b9,b10,b11,b12,b13,b14,b15,b16} have addressed this challenge. \cite{b1} proposed a framework for text transmission that maximizes semantic similarity by jointly optimizing the allocation of resource blocks and semantic feature selection. In \cite{b4}, the authors introduced a method to optimize semantic spectral efficiency by adjusting the number of transmitted semantic symbols and channel assignment. \cite{b5} proposed a method to enhance system throughput through user association and bandwidth allocation. In \cite{b6}, semantic accuracy and conventional constraints were considered to maximize energy efficiency, focusing on power allocation and determining the semantic compression rate. \cite{b7} proposed a controllable compression module that balances content quality and the size of transmitted semantic information. \cite{b8} introduced a method to maximize the effective semantic transmission rate by jointly optimizing the semantic extraction ratio and base station (BS) association while meeting semantic accuracy requirements. Other research has focused on enhancing user experience quality (QoE) \cite{b9}, \cite{b10}, \cite{b11}, improving energy efficiency (minimizing energy consumption) \cite{b12}, and increasing task execution success probability \cite{b13}, \cite{b14}, \cite{b15}. In task-oriented SemCom \cite{b5,b12,b16}, the amount of transmitted data is lower than in full data reconstruction, leading to reduced resource requirements, differentiating it from approaches focused on complete reconstruction. 

Previous works have also adopted simplifying assumptions in resource allocation. A common approach is to assign a single channel per user, as in \cite{b1,b4,b5}, \cite{b12}. Fixed power allocation is another typical assumption \cite{b1}, and \cite{b9}. In contrast, a key advantage of semantic communication lies in its ability to prioritize information based on its relevance to task success. This allows the system to transmit only essential content—determined by the compression rate—while accounting for constraints such as acceptable delay and available spectral or computational resources.

Contrary to prior art that typically allocate identical resources to all users, which often leads to inefficiencies by over-allocating to less critical information and limiting more important content, our method allocates resources based on semantic relevance. Furthermore, strict constraints like semantic similarity reduce network flexibility in meeting user needs and allocating resources. In response, we propose a flexible resource allocation approach based on each user's acceptable semantic similarity range.
\textbf{\textit{The main contributions of this paper are as follows:}}
\begin{itemize} \item We propose a flexible resource allocation framework for multi-user semantic-oriented communication (SOC), which jointly considers required semantic similarity, minimum delay, and SNR. \item We introduce semantic similarity ranges to enhance allocation flexibility, enabling the system to adapt to varying resource availability and user density. \item We formulate the joint optimization of semantic and transmission reliability as a mixed-integer nonlinear programming problem (MINLP). We develop an algorithm to tackle the NP-hard nature of the MINLP problem by iteratively allocating bandwidth and power, while determining appropriate compression rates for each user. \item Simulation results show that our method improves user satisfaction and total network semantic similarity by $17.1\%$ and $14\%$, respectively, compared to rigid semantic similarity-based schemes. \end{itemize}

The remainder of the paper is structured as follows. Section II presents the system model and formulates the resource allocation problem. Section III details the solution approach. Section IV discusses the simulation results and Section V concludes the paper.

\section{System model and problem formulation}
We consider the uplink of a wireless network, where a base station (BS) is located at the center of the coverage area, and $N$ users are uniformly distributed throughout the region. All users and the BS are equipped with a single antenna and are stationary. A centralized resource allocation approach is employed, where the BS assigns resources (power and available bandwidth) to users and determines their compression rates for transmitting semantic information. We assume that all users have similar priority for data transmission, although the expected quality of semantic reconstruction may vary across users.

Deep neural networks used in SemCom for extracting semantic data often need updating when tasks change or when multiple models are required for different tasks. To solve this, \cite{b17} introduced the UDeepSC model, a unified end-to-end framework that supports various tasks and data types. In this study, we adopt UDeepSC for the semantic encoder and decoder. Users employ both a semantic encoder and a channel encoder, while the BS uses a semantic decoder and a channel decoder. This paper focuses on image transmission, where users extract semantic information from images using the semantic encoder. Based on a set compression rate ($O$), the encoded data is transmitted to the BS, which reconstructs the image using both decoders. For simplicity, we assume orthogonal resource allocation (no interference) and an additive white Gaussian noise (AWGN) channel. The received SNR for the $i^{th}$ user is derived as follows:
\begin{equation}
SNR_i=\frac{P_ih_i}{\beta_i N_0}.
\end{equation}

Here, $N_0$ represents the noise power spectral density, $P_i$ denotes the transmission power, and $\beta_i$ is the allocated bandwidth for the $i^{th}$ user. $h_i$ is the channel gain between the $i^{th}$ user and the BS. 
The next section will explain the resource allocation problem in detail.
\subsection{Problem formulation}
As previously mentioned, this paper focuses on full data reconstruction at the receiver, as required in SOC, which requires transmitting a large amount of semantic information over limited spectrum resources. As a result, when many users request service, the BS may not be able to meet all their demands. Therefore, our goal is to enhance the number of satisfied users. Our method addresses both conventional and semantic requirements. Specifically, we propose a resource allocation approach that ensures both transmission and semantic reliability, aiming to maximize the number of satisfied users within the network.

\subsubsection{Reliable Transmission} To ensure transmission reliability, data must be transmitted with high fidelity through the physical channel. This requires an SNR above a certain threshold to minimize bit error rates (BER) and guarantee reliable data recovery at the receiver. Therefore, the following constraint is incorporated into the resource allocation problem:
\begin{align}
0<SNR_{ith}\leq SNR_i.
\end{align}
Here, $SNR_{ith}$ represents the minimum required SNR of the $i^{th}$ user.

\subsubsection{Semantic reliability} Ensuring semantic reliability ($\xi$) requires that the data recovered by the semantic decoder closely aligns with the original data, maintaining an acceptable level of semantic similarity for users. Semantic similarity is a function of $O$ and SNR, denoted as $\xi(O,SNR)$. Higher compression rates increase the demand for computing resources and reduce semantic similarity, while lower rates increase the need for spectrum resources and enhance semantic similarity. Thus, given the trade-off between computational and communication resources in SemCom, a deficiency in one can be compensated by leveraging the other.

Previous research on resource allocation for SemCom typically establishes a fixed threshold for the minimum required semantic similarity. Enforcing a strict semantic similarity constraint requires a specific amount of communication resources, which limits flexibility in resource allocation. To address this, we propose a range for semantic similarity. User $i$ specifies a minimum semantic similarity threshold ($\xi_{i,th_{min}}$), indicating acceptable data retrieval and reliable transmission at the semantic level. Additionally, a maximum semantic similarity threshold ($\xi_{i,th_{max}}$) is defined, reflecting the user's highest satisfaction regarding computational resources and energy consumption. Based on this range, the BS can flexibly adjust the compression rate, transmission power, and assigned bandwidth to serve more users. Consequently, the following constraint is incorporated into the resource allocation problem:

\vspace{-2mm}\begin{equation}
\begin{aligned}
0<\xi_{i,th_{min}}\leq \xi_i(O_i,SNR_i) \leq \xi_{i,th_{max}}\leq 1,
\end{aligned}
\vspace{-0mm}\end{equation}

where $\xi_i(O_i,SNR_i)$ represents the semantic similarity of the $i^{th}$ user. For simplicity, we denote it as $\xi_i$ hereafter.

Furthermore, our proposed method incorporates a delay constraint, ensuring that within a given time interval ($\tau_{ith}$), the required semantic information for user data reconstruction is successfully transmitted. The transmission delay is influenced by the compression rate and power allocation factor. Given the transmission rate of the $i^{th}$ user $R_i=\beta_ilog (1+SNR_i)$ and the raw data size $d_{0,i}$, the transmission time is defined as: $t_i= \frac {d_{0,i}\cdot(1-O_i)}{R_i}$. Thus, the following constraint is imposed on the transmission delay: $t_i\leq \tau_{ith}$.

In our proposed method, transmission is considered reliable, and the corresponding user is deemed satisfied if both SNR and semantic similarity exceed specified thresholds. If $\frac{SNR_{ith}}{SNR_i}$ is less than one, it ensures the transmission reliability for the user $i^{th}$. Similarly, this principle applies to semantic reliability, represented by ($\frac{\xi_{i,th_{min}}}{\xi_i}$). Therefore, if both ratios are less than one, the user is considered satisfied. 
To maximize the number of satisfied users, our objective is to minimize the product of semantic and transmission reliabilties across all users, which is defined as:
$\sum\limits_{n=1}^{N} (\frac{SNR_{ith}}{SNR_i})^ {a} \cdot (\frac{\xi_{i,th_{min}}}{\xi_i})^{a}$.
Here, $a$ represents a penalty factor and selecting a large $a$ penalizes unsatisfied users. This implies that if any of the ratios exceed one, the objective function increases. Furthermore, since $R_i$ represents the transmission rate of the $i^{th}$ user, multiplying SNR by semantic similarity yields the effective transmission rate. This suggests that even if a user's SNR is high, a low level of semantic similarity can result in resources being allocated to transmit less critical information. Consequently, resource efficiency decreases, and the value of (5) is reduced. 

The overarching goal is to maximize the number of satisfied users by minimizing the product of semantic and transmission reliability metrics, thereby enhancing network capacity while meeting users' requirements. The detailed problem formulation is presented below.

\vspace{-2mm}\begin{align}
& F:\underset{\{\beta_n,O_n,P_n\}_{\forall n}}{\text {min}}\sum_{n=1}^{N} (\frac{SNR_{ith}}{SNR_i})^ {a} \cdot (\frac{\xi_{i,th_{min}}}{\xi_i})^{a} \notag\\ 
& \text{subject to:} \notag\\
& C_1: 0<\sum_{n=1}^{N} \beta_i\leq M,\;\;\;\; i\in N, \notag\\
& C_2: \beta_{i,min} \leq \beta_i,\;\;\;\; i\in N, \notag\\
& C_3: 0<P_{i} \leq P_{tot},\;\;\;\; i\in N, \notag\\
& C_4: 0<SNR_{ith} \leq SNR_i,\;\;\;\; i\in N, \notag\\
& C_5: 0<O_i \leq 1,\;\;\;\; i\in N, \notag\\
& C_6: t_i \leq \tau_{ith},\;\;\;\; i\in N, \notag\\
& C_7: 0<\xi_{i,th_{min}}\leq \xi_i \leq \xi_{i,th_{max}}\leq 1,i\in N.
\label{eq_F}
\end{align}

Here, $n=\{1,\cdots,N\}$ is the user set, and $\beta_i$ denotes the allocated bandwidth for user $i$. $C_1$ ensures that the total allocated bandwidth does not exceed the available bandwidth ($M$).
 $C_2$ ensures that each user’s minimum bandwidth requirement is satisfied. $C_3$ guarantees that the transmission power ($P_i$) for each user remains within the maximum allowed power ($P_{tot}$), while $C_4$ ensures that each user meets their minimum required SNR ($SNR_{ith}$) for reliable transmission. The compression rate directly impacts semantic similarity and is determined for each user based on factors like semantic similarity range and available bandwidth. The range for the compression rate $O_i$ is $[0,1]$, as ensured by $C_5$. $C_6$ ensures that the user transmission delay remains within acceptable limits. Finally, $C_7$ ensures that the users meet the semantic similarity constraint. Next, we focus on optimizing $\beta_i$, $O_i$, and $P_i$ by minimizing the product of transmission and semantic similarity metrics, subject to the system constraints defined in \eqref{eq_F}.

\vspace{-1mm}\section{Proposed resource allocation algorithm}
\vspace{-1mm}

Problem~\eqref{eq_F} is a non-convex nonlinear problem and belongs to the class of MINLP problems. Generally, MINLP problems are NP-hard (Non-deterministic Polynomial-time hard) and cannot be efficiently solved using traditional optimization methods. Furthermore, there is no closed-form expression for the semantic similarity indicator ($\xi_i$). Therefore, due to the interdependencies between semantic similarity, compression rate, and minimum acceptable delay, we decompose ~\eqref{eq_F} into two sub-problems of power and bandwidth allocation ($F_1$) and compression rate adjustment ($F_2$), formulated as follows:
\begin{equation}
\vspace{-4mm}\begin{aligned}
\vspace{-2mm}& F_1:\underset{\beta,P}{\text {min}}\sum_{n=1}^{N} (\frac{SNR_{ith}}{SNR_i})^ {a} \cdot (\frac{\xi_{i,th_{min}}}{\xi_i})^{a} \notag\\ 
& \text{subject to:\,} 
C_1,C_2,C_3,C_4.
\vspace{-2mm}\end{aligned}
\end{equation}
and
\begin{align}
& F_2:\underset{O}{\text {min}}\sum_{n=1}^{N} (\frac{SNR_{ith}}{SNR_i})^ {a} \cdot (\frac{\xi_{i,th_{min}}}{\xi_i})^{a} \notag\\ 
& \text{subject to:\,} 
C_5,C_6,C_7.
\end{align}

Based on the objective function and constraints of $F_1$, the problem can be formulated as a geometric programming (GP) problem, solvable with tools like CVX.

By substituting (1) into $F_1$ and performing some manipulations, (7) can be rewritten as (9). 
\begin{align}
& F_1^{'}:\underset{\beta,P}{\text {min}}\sum_{n=1}^{N} (SNR_{ith})^ {a} \cdot (\xi_{i,th_{min}})^ {a} \cdot (P_i)^ {-a} \cdot (h_i)^ {-a} \notag\\
&\cdot (\beta_i)^ {a} \cdot (\xi_i)^ {-a} \cdot (N_0)^ {a} \notag\\ 
& \text{subject to:} \notag\\
& C_1^{'}: 0<\sum_{n=1}^{N} M^{-1} \cdot \beta_i \leq 1 \notag\\
& C_2^{'}: 0<  \beta_i^{-1}\cdot \beta_{i,min} \leq1 \notag\\
& C_3^{'}: 0< P_{tot}^ {-1}\cdot P_{i} \leq 1 \notag\\
& C_4^{'}: 0< SNR_{ith}\cdot P_{i}^{-1}\cdot h_{i}^{-1} \cdot \beta_i \cdot N_0 \leq 1,
\vspace{-2mm}\end{align}

Since there is no closed-form expression for semantic similarity, we run the selected encoder-decoder model (UDeepSC) over specific SNR and compression rate ranges with defined step sizes to obtain similarity values. The resulting similarity table is then used to solve $F_2$. To address problem (6), we propose Algorithm 1, which iteratively solves $F_2$ and $F_1^{'}$ to reach a suboptimal solution. 

To solve $F_2$ , the users’ transmission reliability is prioritized in the first step. For each user, based on the iteration’s SNR ($SNR_{i, it}$), initialized with $SNR_{ith}$, and $C_7$, the best similarity value $\xi_i$ is selected by searching the lookup table. The nearest similarity value less than $\xi_{i,th_{max}}$ and satisfying $C_7$ is selected as optimal. With the obtained $SNR_{i,it}$, and $\xi_i$, the corresponding compression rate from the lookup table is selected as the solution to $F_2$ (i.e. $O_i$). If selected $\xi_i$ and $O_i$ satisfy $C_6$, they are used to solve $F_1^{'}$. Otherwise, the next best similarity value ($\xi_i$) is selected based on $SNR_{i,it}$, iterating until no further options exist. In such a case, $\xi_i$ and $O_i$ are obtained from the lookup table using $SNR_{ith}$. 

\begin{figure}[t]
\vspace{-4mm}
\begin{algorithm}[H]
\caption {Proposed algorithm for bandwidth, power allocation, and compression rate adjustment.}\label{alg:one}
\textbf{Input: }{$N, M, P_{tot}, a, N_0,$ $H=\{h_1,…,h_N\}$

$SNR_{ith}=\{SNR_{1th},...,SNR_{Nth}\}$,

$\xi_{i,th_{min}}=\{\xi_{1,th_{min}},...,\xi_{N,th_{min}}\}$,

$\xi_{i,th_{max}}=\{\xi_{1,th_{max}},...,\xi_{N,th_{max}}\}$
}

\KwResult{$\beta,O,P$}
 initialization\; 
 Set $SNR_{i,it}=SNR_{ith}$ for each user, $i$.
 
 Set $iter = 0$
 
 Set $\beta_i = \beta_{i,min}$

\Repeat {$iter< iter_{max}$  or $\Delta F<\partial$ }{
 \For{$i = 1$ to $N$}{ 
    \While{true}{
          With the use of a similarity table, find:
          Next best optimal $\xi_i$  and its corresponding $O_i$ for $SNR_{i,it}$ satisfying $C_7$
          
          \eIf{$C_6$ is satisfied}
           {
             \textbf{break\;}
           }{
           \If{there is no more optimal 
             $\xi_i$ candidate}
           {
             Set $\xi_i$ to the best similarity corresponding to the $SNR_{ith}$ using the similarity table
             
             Set $O_i$ to the corresponding compression rate of selected $\xi_i$ and $SNR_{ith}$ from the similarity table
            
            \textbf{break\;}
           }
          }
    }
    
 }

    Solve $F_1^{'}$ as a GP problem using each user’s $\xi_i$
    
    Calculate each $SNR_{i,it}$ using calculated $\beta_i$, $P_i$
    
    Set $iter=iter+1$
}
\end{algorithm}
\end{figure}

\textbf{\emph{Computational Complexity:}} The complexity of the proposed algorithm is influenced by several factors: the number of iterations ($it$), the lookup table method, and the complexity of the GP problem. Since $\xi_i$ can only be obtained through the lookup table method, an exhaustive search method is used. The complexity of this search is acceptable due to the limited ranges for SNR and compression rate. Therefore, the complexity of it is $O(kN)$, where $N$ is the number of users and $k$ is the size of the similarity table.  Additionally, the convexity and duality properties of GP are well understood, and robust numerical solvers for GP are readily available. In practice, interior point methods applied to GP exhibit polynomial time complexity ($O(n^3)$, where $n$ is the number of variables and constants.) and perform efficiently with high-quality software (e.g., the MOSEK package) \cite{b18}, \cite{b19}. Therefore, considering the polynomial complexity of GP and the limited number of iterations in the algorithm, the overall complexity of the problem is polynomial in terms of the number of users ($O(it×(kN+(7N+3)^{3} ))$).

\section{Evaluation and simulation results}\vspace{-1mm}
The performance of the proposed method is evaluated through a simulation study, as detailed in this section. In our simulation, we use Python to run the UDeepSC model, which was implemented using PyTorch and trained on the CIFAR-10 dataset. Additionally, MATLAB is employed to simulate the proposed algorithm and solve the GP problem using CVX. We compare the proposed method with three other approaches: the baseline method, where users transmit data using conventional communication, the strict similarity method (our method without similarity rang), where each user has a predefined minimum semantic similarity and the proposed method in \cite{b11} (QoE-based RA). The QoE-based resource allocation (RA) method employs a strict semantic similarity constraint and assigns a single channel to each user. Its objective is to maximize both the semantic rate and semantic similarity in alignment with users' preferences. The following metrics are evaluated:
\begin{itemize}
\item Number of satisfied users: Users are considered satisfied when both the minimum semantic similarity ($\xi_{i,th_{min}}$) and minimum SNR ($SNR_{ith}$) are met.
\item Semantic similarity of the users: Semantic similarity measures how closely the reconstructed image matches the original data. Here, Peak Signal-to-Noise Ratio (PSNR) is used to assess image reconstruction performance by calculating the ratio between the maximum possible power and noise.
\end{itemize}
\begin{equation}
PSNR=10log_{10}(\frac{P_{max}^{2}}{MSE})
\end{equation}
\begin{equation}
MSE=\frac{1}{n}\sum_{n=1}^{n}(SI_i-RI_i)^2,
\vspace{-2mm}\end{equation}

where MSE represents the mean-square error between the source image and reconstructed image, $P_{max}$ is the maximum pixel value (255), and $n$ is the total pixel count.
Simulations are conducted in a cell with a 100m radius and uniformly distributed users. The values of  $\frac{SNR_{ith}}{SNR_i}$  and $\frac{\xi_{i,th_{min}}}{\xi_i}$   in equation (6) range from 0 to 1 for the satisfied users. To reflect the effect of unsatisfied users, we set a=2, although larger exponents could also be used. In the proposed method, $\xi_{i,th_{min}}$ and $\xi_{i,th_{max}}$ selected within the  range of $0.6$ to $0.9$. In the strict similarity method, the minimum semantic similarity for each user is set within this range. Other parameters are listed in Table I. 

Figure 1 shows the number of satisfied users as a function of the maximum available bandwidth. In the proposed method, users have a range for semantic similarity, allowing the BS to allocate resources more flexibly. When the BS cannot serve all users with the available resources, it adjusts the compression rate for each user based on their semantic similarity range to serve more users. This flexibility enables the proposed method to increase the number of satisfied users by dynamically adjusting the acceptable semantic similarity and compression rate. In contrast, the strict similarity method allocates resources to ensure users' strict requirements are met. As a result, users must transmit specific amounts of semantic information to guarantee semantic similarity, leading to higher resource consumption and fewer satisfied users compared to the proposed method. Additionally, the baseline method, which uses classical communication, results in the fewest satisfied users because it transmits raw data, consuming more resources. Moreover, the QoE-based RA method results in fewer satisfied users compared to the proposed method. This limitation arises from its exclusive allocation of a single channel per user, even when additional channels are available. Consequently, users are deprived of these resources, leading to a reduction in transmitted semantic features, and ultimately, decreased semantic similarity and satisfaction. Furthermore, if the requirements of a user are not met, it results in resource wastage, prevents other users from accessing those resources, and ultimately reduces the overall number of satisfied users. In addition, this method also has less flexibility in resource allocation and user satisfaction level due to the use of strict semantic similarity constraints. Additionally, since the proposed method serves a higher number of users with the same bandwidth, its resource efficiency in terms of the number of satisfied users is greater than that of the other methods. 

Figure 2 shows the semantic similarity values provided to satisfied users. In this figure, the semantic similarity ranges for users in the proposed method are displayed next to the corresponding bar chart, while the minimum semantic similarity for users in the strict similarity method and QoE-based RA are shown above the respective bar chart. In the proposed method, all users, except for user 8, were served within their semantic similarity range. However, in the strict similarity method, users 5, 8, and 10 were not served due to the rigid semantic similarity requirements, resulting in more resources being allocated to the users who were served. In contrast, the proposed method improves resource allocation by reducing users' resource consumption and providing a semantic similarity slightly below their maximum request ($\xi_{i,th_{max}}$). This approach allows the proposed method to satisfy more users while ensuring the provided similarity remains above the minimum required value ($\xi_{i,th_{min}}$), thus keeping users satisfied. 

Furthermore, the QoE-based RA method, due to its strict semantic similarity constraint and the allocation of a single channel per user, can lead to user dissatisfaction when the similarity constraints are not met. This single-channel allocation also results in the underutilization of channels assigned to dissatisfied users, preventing other users from accessing these channels. Consequently, the number of satisfied users is lower compared to the other two methods. 

Figure 3 shows the average semantic similarity of the network as a function of the maximum available bandwidth. As seen, the proposed method has lower average similarity compared to the strict similarity method. This decrease is intentional, as the proposed method prioritizes serving more users. The trade-off between lower similarity and accommodating more users is acceptable. On the other hand, the strict similarity method uses a more rigid constraint, resulting in higher similarity but fewer users being served, leading to a higher average similarity than the proposed method. While the QoE-based RA method incorporates a strict semantic similarity constraint, its average semantic similarity is lower than that of the strict semantic similarity method due to the allocation of a single channel per user and the underutilization of channels assigned to dissatisfied users. These factors also make its average semantic similarity ($0.7058$) almost the same as that of the proposed method ($0.7085$).
\begin{table}[t]
\vspace{0.05in}\caption{SIMULATION PARAMETERS}\vspace{-4mm}
\begin{center}
\begin{tabular}{|c|c|}
\hline
\textbf{Parameter} & \textbf{\textit{Value}} \\
\hline
\text{Maximum bandwidth} & \text{\textit{8-25 MHZ}} \\
\hline
\text{User’s Max. transmission power} & \text{\textit{0.5W}} \\
\hline
\text{Noise power spectral density, $N_0$} & \text{\textit{-173dBm/Hz}} \\
\hline
\text{Users’ delay constraint } & \text{\textit{0.4-0.6 ms}} \\\hline
\text{Initial data size, $d_0$ } & \text{\textit{3-5 Mbit}} \\\hline
\text{Minimum Users SNR} & \text{\textit{20-25dB}} \\
\hline
\end{tabular}
\label{tab1}
\end{center}
\vspace{-4mm}\end{table}

\begin{figure}[t]
\centerline{\includegraphics[width=1.05\linewidth, height=5cm]{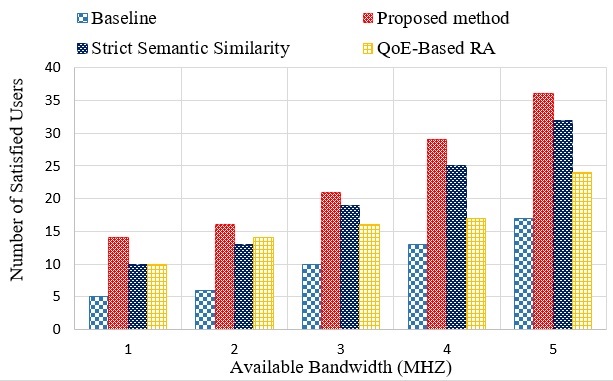}}
\vspace{-2mm}\caption{Number of satisfied users VS maximum available bandwidth.}
\label{fig}\vspace{-1mm}
\vspace{-2mm}\end{figure}

\begin{figure}[t]
\centerline{\includegraphics[width=1.05\linewidth, height=5cm]{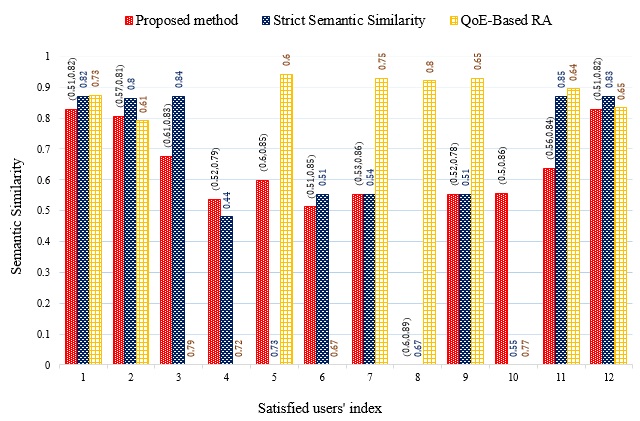}}
\caption{Semantic similarity of the satisfied users.}
\label{fig}
\end{figure}

\begin{figure}[t]
\centerline{\includegraphics[width=1.05\linewidth, height=4.5cm]{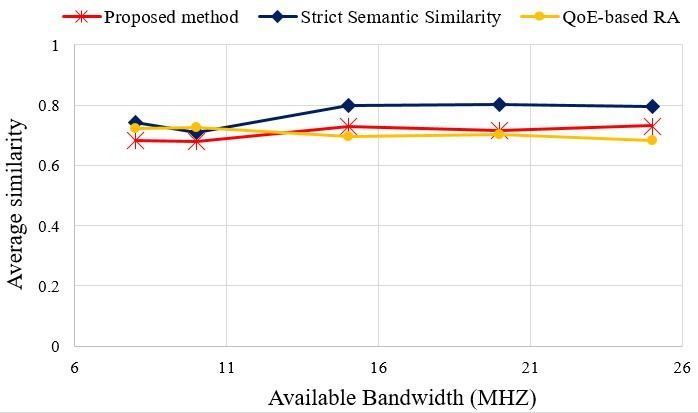}}
\vspace{-3mm}\caption{Average similarity VS maximum available bandwidth.}
\label{fig}
\vspace{-2mm}\end{figure}

\vspace{-3mm}\section{Conclusion}
This paper proposed a flexible resource allocation method for SOC. Our algorithm optimizes resource allocation in SemCom by dynamically adjusting the compression rate, transmission power, and bandwidth to ensure that users' data retrieval meets a specified range of semantic similarity thresholds. The algorithm incorporates a minimum and maximum semantic similarity threshold for each user, allowing flexibility in resource allocation while maintaining acceptable data quality. Additionally, a delay constraint ensures the timely transmission of the required semantic information for data reconstruction. Simulation results show that a slight, yet acceptable, reduction in semantic similarity for some users enables more efficient resource allocation, allowing the system to serve more users while maintaining overall satisfaction within the network. For future work, we will extend the proposed method to accommodate various data types and integrate task- and goal-oriented semantic communications into the resource allocation process.

\end{document}